# Correct quantitative determination of ethanol and volatile compounds in alcohol products

**Siarhei Charapitsa[a] Nikita Kulevich[a] Svetlana Sytova[a] Yurii Yakuba[b]**

[a] *Research Institute for Nuclear Problems of Belarusian State University, Bobruiskaya Str. 11, Minsk 220030, Belarus*

[b] *North-Caucasian Zonal Research Institute of Horticulture and Viticulture, 40 Let Pobedy Str. 39, Krasnodar 350901 , Russia*

Keywords: alcohol products; ethanol concentration; volatile compounds.

## 1. INTRODUCTION

The standards for quality and safety control of alcohol production [1-2] prescribe determination of the following volatile compounds: acetaldehyde, methyl acetate, ethyl acetate, methanol, 2-propanol, 1-propanol, isobutyl alcohol, n-butanol, izoamyl alcohol. Results of the analysis are expressed in milligrams per litre (mg/L) of absolute alcohol (AA). Such analysis is carried out by the Internal Standard (IS) method. 1-pentanol and 2-pentanol are most commonly used as IS. This method ensures high data reliability. However, the procedure of introducing of an internal standard substance in the sample at the level of some ppm requires a high level of laboratory technicians, performing analyses. It was proposed [3] to use ethanol as IS for the analysis of alcohol production. Analysis of alcohol production in this case consists in the traditional procedure of determining the relative ratios of the detector response (*Relative Response Factors – RRF*) of analysed impurities with respect to ethanol by standard solutions and then the subsequent use of these coefficients in the calculation of concentration of impurities. It should be noted that for modern chromatographs coefficients *RRF* are enough stable and can be tabulated.

## 2. MATERIALS AND METHODS

Analysis on validation of our method was carried out in the Laboratory of Analytical Research of Research Institute for Nuclear Problems of Belarusian State University. We use gas chromatograph Crystal-5000 equipped with FID. All individual standard compounds were purchased from Sigma-Fluka-Aldrich (Germany). The standard solutions for calibration and sample solutions were prepared by adding the individual standard compounds to the ethanol-water mixture (96:4) by gravimetric method. A typical chromatogram of the used solutions is presented in Figure 1. To show the dominant ethanol compound and other minor compounds simultaneously the logarithmic scale of the response signal was chosen.

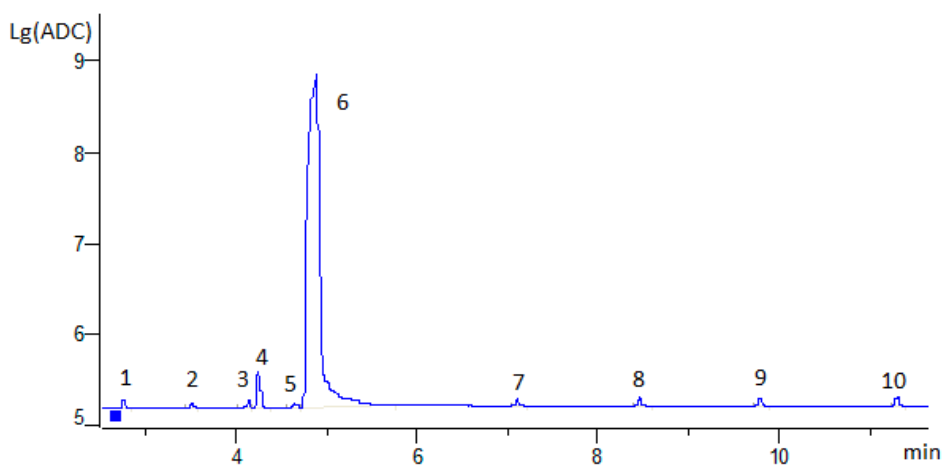

Figure 1. Typical chromatogram of the standard ethanol-water solution (96% and 4%). 1 – acetaldehyde, 2 – methyl acetate, 3 – ethyl acetate, 4 – methanol, 5 – 2-propanol, 6 – ethanol, 7 – 1-propanol, 8 – isobutyl alcohol, 9 – n-butanol, 10 – isoamyl alcohol.

## 3. RESULTS AND DISCUSSION

To demonstrate the reliability of the proposed method the standard ethanol-water (96:4) solution with initial volatile compounds concentration about 4000 mg/L(AA) was analyzed after dilution with water in the ratios 1:1, 1:9, 1:99, 1:1999 and 1:9999. Experimental results are presented in Table 1. Example of experimental data for methanol is presented in Figure 2. Here line 1 is the detector response versus the amount of compound. Lines 2 and 3 are the detector response versus the concentration of compound in mg/L(sol) and in mg/L(AA), respectively. We have similar plots for remaining compounds.

Even after dilution with water in the ratio 1:999, the difference between the measured concentrations of all compounds and their values calculated using the gravimetric method does not exceed 7.7 %. With the dilution 1:9999 there are peaks of methanol and ethanol only. Other compounds are significantly less than the level of detection. It should be noted the relative discrepancy of measured concentrations of methanol does not exceed 6.6%.

Table 1. The measured concentrations of analyzed volatile compounds and ethanol, presented according to the degree of dilution with water.

| | Measured concentration mg /L (AA) (Relative standard deviation,%) [Concentration under certificate (mg /L (AA)) / (mg /L (sol))] | | | | | | | | | |
|---|---|---|---|---|---|---|---|---|---|---|
| | Compound | | | | | | | | | |
| Dilution | acetaldehyde | methyl acetate | ethyl acetate | methanol | 2-propanol | ethanol | 1-propanol | isobutyl alcohol | n-butanol | isoamyl alcohol |
| - | 4556 (6,6) [4275/3 768] | 4436 (0,9) [4397 / 3875] | 4253 (1,9) [4173 / 3678] | 42586 (1,4) [41995 / 37017] | 4112 (3,0) [3991 / 3518] | N/A [789300 / 695748] | 4076 (1,6) [4012 / 3 536] | 4049 (1,9) [3975 / 3504] | 4174 (2,5) [4071 / 3588] | 4458 (9,5) [4071 / 3588] |
| 1:1 | 4451 (4,1) [4275 / 1884] | 4127 (-6,1) [4397 / 1938] | 4018 (-3,7) [4173 / 1839] | 40462 (-3,7) [41995 / 18509] | 4000 (0,2) [3991 / 1759] | N/A [789300 / 347874] | 3973 (-1,0) [4012 / 1768] | 4007 (0,8) [3975 / 1752] | 4096 (0,6) [4071 / 1794] | 4412 (8,4) [4071 / 1794] |
| 1:9 | 4340 (1,5) [4275 / 377] | 3961 (-9,9) [4397 / 388] | 3780 (-9,4) [4173 / 368] | 39043 (-7,0) [41995 / 3702] | 3875 (-2,9) [3991 / 352] | N/A [789300 / 69575] | 3868 (-3,6) [4012 / 354] | 3904 (-1,8) [3975 / 350] | 4012 (-1,4) [4071 / 359] | 4318 (6,1) [4071 / 359] |
| 1:99 | 4406 (3,1) [4275 / 37,7] | 4002 (-9,0) [4397 / 38,8] | 3762 (-9,8) [4173 / 36,8] | 38645 (-8,0) [41995 / 370,2] | 3866 (-3,1) [3991 / 35,2] | N/A [789300 / 6958] | 3862 (-3,7) [4012 / 35,4] | 3903 (-1,8) [3975 / 35,0] | 4107 (0,9) [4071 / 35,9] | 4479 (10,0) [4071 / 35,9] |
| 1:999 | 4280 (0,1) [4275 / 3,77] | 4292 (-2,4) [4397 / 3,88] | 4107 (-1,6) [4173 / 3,68] | 38764 (-7,7) [41995 / 37,02] | 3818 (-4,3) [3991 / 3,52] | N/A [789300 / 696] | 3820 (-4,8) [4012 / 3,54] | 4140 (4,1) [3975 / 3,50] | 4024 (-1,2) [4071 / 3,59] | 3937 (-3,3) [4071 / 3,59] |
| 1:9999 | N/A | N/A | N/A | 39210 (-6,6) [41995 / 3,702] | N/A | N/A [789300 / 69,6] | N/A | N/A | N/A | N/A |

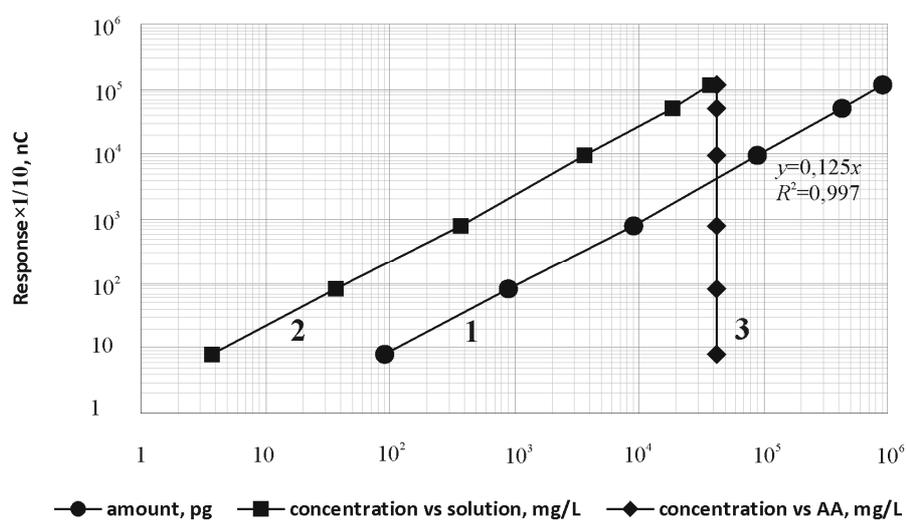

Figure 2. Experimental data for methanol.

In Fig.2 for line 1 corresponding dependence of type $y=ax$ is given. The magnitude of the correlation coefficient $R^2$ was 0,997. For all compounds we have $R^2$ in the range between 0,996 and 0,999.

Our proposed method is original and innovative. It improves the reliability of the measured data as well as substantially simplifies the whole measurement procedure. The "Method of measurement of the mass concentration of volatile compounds in alcohol drinks by gas chromatography" has been validated by the Federal Agency for Technical Regulation and Metrology of Russian Federation (*Rosstandart*) in July, 2013 (Certificate No. 253.0169/01.00258/2013). We hope to start mutually beneficial collaboration in including of the new method of quantitative determination of volatile compounds in alcohol products to the measurement practices of OIV.

The proposed method can be easily incorporated into daily practice of analytical and control laboratories because for its implementation there are no additional material, financial or time costs.

**References.**

1. International Organization of Vine and Wine (OIV). Compendium of international methods of wine and must analysis, 2009; Vols. 1 and 2.
2. Commission Regulation (EC) No 2870/2000 of 19 December 2000 laying down Community reference methods for the analysis of spirits drinks, 2000.
3. S.V. Charapitsa et al. J. Agric. Food Chem. 61(2013)2950.